\documentclass[english,english,pra,english,preprint,amsmath,amssymb,aps,longbibliography,showkeys, titlepage]{revtex4-2}
\usepackage{lmodern}
\usepackage[T1]{fontenc}
\usepackage[latin9]{luainputenc}
\usepackage{amsmath}
\usepackage{amsthm}
\usepackage{amssymb}
\usepackage{graphicx}

\makeatletter
\usepackage[T1]{fontenc}
\usepackage{amsmath}
\usepackage{amsthm}
\usepackage{amssymb}
\usepackage{graphicx}

\usepackage{graphicx}
\usepackage{float}
\usepackage{xcolor}
\usepackage[hypertexnames=false]{hyperref}
\hypersetup{
	breaklinks = true,
    colorlinks = true,
    citecolor = {blue},
	urlcolor = {blue},
	linkcolor = {blue}
}

\ifdefined\showcaptionsetup
 \PassOptionsToPackage{caption=false}{subfig}
\fi
\usepackage{subfig}
\makeatother

\usepackage{babel}
\begin{document}
\title{Second quantization of nonlinear Vlasov--Poisson system for quantum
computation}
\author{Michael Q. May}
\email{mqmay@princeton.edu}

\affiliation{Princeton Plasma Physics Laboratory, Princeton University, Princeton,
NJ 08540}
\affiliation{Department of Astrophysical Sciences, Princeton University, Princeton,
NJ 08540}
\author{Hong Qin}
\email{hongqin@princeton.edu }

\affiliation{Princeton Plasma Physics Laboratory, Princeton University, Princeton,
NJ 08540}
\affiliation{Department of Astrophysical Sciences, Princeton University, Princeton,
NJ 08540}
\begin{abstract}
The Vlasov--Poisson equations, fundamental in plasma physics and
astrophysical applications, are rendered linear, finite-dimensional,
and discrete by second quantization. Conditions for correspondence
between the pre-quantized and quantized equations are derived, and
numerical simulations demonstrating the quantized linear system can
capture nonlinear dynamics are presented. Finally, encouraging scaling
relations emphasizing the prospect of using quantum computers to efficiently
integrate the second quantized Vlasov--Poisson equations as a model
for the usual Vlasov--Poisson equations are derived. 
\end{abstract}
\maketitle

\section{Introduction}

Existing quantum algorithms for Hamiltonian dynamics require finite-dimensional,
linear, and unitary operators, but the Vlasov--Poisson equations,
foundational in plasma physics and gravitational dynamics, are instead
infinite-dimensional, nonlinear, and symplectic. Efficient quantum
algorithms have been developed to simulate linear Vlasov--Poisson
dynamics \citep{engel19,novikau22,ameri23,toyoizumi24,miyamoto24},
but simulation of the nonlinear problem has remained elusive. General
approaches to nonlinearity for quantum computing have been proposed
on multiple fronts, which may be divided into three broad categories.
First, one can simulate multiple copies of the quantum system and
destroy half the copies at each nonlinear step \citep{lloyd20}. This
has the obvious drawback of requiring the initial simulation of $2^{n}$
copies of the system, where $n$ is the number of nonlinear steps.
Second, one can employ quantum variational algorithms \citep{mocz21,cappelli24},
in which quantum computation is solely used to efficiently evaluate
some linear cost function, or quantum-linear, classical-nonlinear
algorithms \citep{higuchi24}, in which all linear operators occur
on the quantum computer, e.g. particle pushing. This requires an efficient
method of rapidly reading from and writing to a quantum register,
which introduces new failure modes into the calculation. Third, one
can transform the original, nonlinear system into a linear system,
for which many efficient Hamiltonian algorithms already exist \citep{hhl,low17,campbell19}. 

The third approach has the most promise, assuming that a useful linearization
scheme may be found. For specific nonlinear systems with sufficient
dissipation, efficient algorithms exist \citep{liu21,krovi23,xue21}.
These take advantage of Koopman von Neunam (KvN) embedding \citep{koopman31,neumann1932}
or Carleman linearization \citep{carleman32}, in which any nonlinear
system of equations may be embedded into a linear system. KvN linearization
has been also discussed as a more general method for simulation of
nonlinear systems which are not necessarily dissipative \citep{jin23,joseph20}.
In almost all cases, the KvN or Carleman linearizations result in
an infinite dimensional system, and some kind of truncation and closure
must then be applied to render the set of equations finite. The depth
of the truncation and choice of closure may non-trivially affect the
physics one is attempting to capture. 

Alternatively, second quantization may be used to linearize a nonlinear
system for quantum simulation. The method is physically motivated
and no part of the quantization procedure is arbitrary. Rather than
embed the nonlinear system in a linear one, the nonlinear system is
instead transformed to a linear quantum system which becomes the nonlinear
system in the classical limit. Second quantization has previously
been used to render the dynamics of the quantum harmonic oscillator
finite and discrete \citep{adqo} and to simulate the nonlinear three-wave
interaction (TWI) on a quantum computer \citep{shiPhysRevA,Shi2017,Shi2021,Shi2018}.
While second quantization does introduce quantum artifacts, including
dispersion and uncertainty relations, to the system, the method of
addressing these artifacts is simply to increase the resolution of
the simulation. It has been shown that the classical three-wave instability
and nonlinear three-wave oscillation may each be captured in systems
linearized via second quantization \citep{may_qin,may_qin_nonlinear_three_wave}. 

In the next section, we will detail the procedure for second quantizing
the Vlasov--Poisson system. As a preparatory step, the system is
first Schr\"odingerized and decomposed into modes, a process which is
a well-known computational tool for the Vlasov--Poisson system \citep{bertrand80,mocz18,mocz21,qin_may_25}.
We then promote the mode amplitudes to operators and discuss the resulting
linear quantum system's dimension and sparsity, both of which effect
the efficiency of known quantum Hamiltonian algorithms \citep{hhl,low17,toyoizumi24}.
Conditions for correspondence between the pre-quantized, nonlinear
and quantized, linear system are derived. In Section III, simulations
of the three-mode and five-mode quantized, linear systems are presented,
in which the five-mode system is shown to faithfully capture a nonlinearities.
In Section IV, we conclude with a discussion of the results and derive
favorable scalings for the prospect of using second quantization to
prepare the nonlinear Vlasov--Poisson equation for efficient quantum
computation. 

\section{Second Quantization}

The Vlasov--Poisson system, 
\begin{align*}
\partial_{t}f+\mathbf{v}\cdot\partial_{\mathbf{x}}f+\frac{q}{m}\mathbf{E}\cdot\partial_{\mathbf{v}}f_{} & =0,\\
\nabla^{2}\phi_{E}+\frac{\rho}{\varepsilon} & =0
\end{align*}
may be made unitary, linear, and finite via three steps: 1) Schr\"odinger
substitution, 2) mode decomposition, and 3) operator promotion (second
quantization). The first two steps condition the set of equations
for the third step, which linearizes the system. By exchanging the
six-dimensional, real-valued distribution function $f(x,v,t)$ for
the three-dimensional, complex-valued wavefunction $\psi_{p}(x,t)$,
the Vlasov--Poisson equations become the Schr\"odinger--Poisson equations:
\begin{align}
\partial_{t}\psi_{p} & =\frac{-i}{\hbar}\left(-\frac{\hbar^{2}}{2m}\nabla^{2}+q\phi_{E}\right)\psi_{p},\label{eq:pre_psi}\\
\nabla^{2}\phi_{E} & =-4\pi nq\left(\psi_{p}^{*}\psi_{p}-1\right).\label{eq:pre_phi}
\end{align}
Here, a length scale has been introduced and set to $1$ to regularize
the second equation, $n$ is the total number of particles, and $\hbar$
is a small quantum parameter. Note that these Schr\"odinger--Poisson
equations are ultimately meant to refer to a classical system, so
$\hbar$ does not need to take the value of the familiar physical
constant. It only needs to be made small enough to recover the Vlasov--Poisson
dynamics. 

One may transform back to phase space using a Wigner transform, 
\begin{equation}
f_{QP}(x,v,t)=\frac{1}{2\pi\hbar}\int_{0}^{1}\psi_{p}(x+y/2,t)\psi^{*}(x-y/2,t)e^{-ivy/\hbar}dy\label{eq:wig}
\end{equation}
but the resultant distribution function is not guaranteed to be positive
\citep{bertrand80}. The symbol $f_{QP}$ indicates it is a quasi-probability
distribution. Convolving the result with some Gaussian kernel (the
full effect being a Husimi transform), one may recover the classical
result in the classical limit (where $\hbar\rightarrow0$). Aside
from the classical limit, the Schr\"odinger--Poisson equations are
also interesting because they describe superfluid Bose--Einstein
condensates, which may act as a model for dark matter \citep{mocz18,mocz21}. 

In the second step, we decompose the wavefunction into modes,

\begin{equation}
\psi_{p}(x,t)=\sum_{l\in\boldsymbol{l}}a_{l}(t)e^{2\pi ilx},\label{eq:mode_decomp}
\end{equation}
for some finite set of modes $\boldsymbol{l}\in\mathbb{Z}^{\text{dim}(\boldsymbol{l})}$,
and then substitute it into Eqs. (\ref{eq:pre_psi}) and (\ref{eq:pre_phi})
to find a set of equations governing the mode evolution
\begin{equation}
\partial_{t}a_{l}=\frac{\delta}{2}a_{l}(2\pi l)^{2}+\frac{1}{4\pi^{2}\delta}\sum_{p,r\in\boldsymbol{l};r\not=l}(r-l)^{-2}a_{p}^{*}a_{l+p-r}a_{r}.\label{eq:adot}
\end{equation}
The mode amplitude $a_{l}$ has been made dimensionless, and $\delta$
is the dimensionless $\hbar$. These dynamics are governed by the
Hamiltonian 
\begin{equation}
H=\frac{\delta}{2}\sum_{l\in\boldsymbol{l}}a_{l}^{*}a_{l}(2\pi l)^{2}+\frac{1}{8\pi^{2}\delta}\sum_{p,r,l\in\boldsymbol{l};r\not=l}(r-l)^{-2}a_{l}^{*}a_{p}^{*}a_{l-r+p}a_{r}.\label{eq:pre_ham}
\end{equation}

Remarkably, Eqs. (\ref{eq:adot}) and (\ref{eq:pre_ham}) remain Hermitian
for any set of modes $\boldsymbol{l}$, but they are still nonlinear.
To linearize them in the third step, we employ the mechanically simple
process of second quantization, where the mode amplitudes become operators
\begin{equation}
a_{l}\rightarrow\hat{a}_{l},\ \ a_{l}^{*}\rightarrow\hat{a}_{l}^{\dagger},
\end{equation}
acting on a Fock space. Denote the set of basis vectors of the Fock
space in the occupation number representation with set total quanta
$N$ by 
\begin{equation}
\Phi_{N}=\left\{ \boldsymbol{\phi}_{k}=\left(\phi_{k,l_{1}},\phi_{k,l_{2}},\dots\phi_{k,l_{\text{dim}(\boldsymbol{l})}}\right)\bigg|\boldsymbol{\phi}_{k}\in\mathbb{N}^{\text{dim}(\boldsymbol{l})},\sum_{l\in\boldsymbol{l}}\boldsymbol{\phi}_{k,l}=N\right\} .
\end{equation}
The creation and annihilation operators have the usual action on elements
of the Fock space, e.g.
\begin{equation}
\hat{a}_{p}(\phi_{k,l_{1}},\phi_{k,l_{2}},\dots,\phi_{k,p},\dots,\phi_{k,l_{\text{dim}(\boldsymbol{l})}})=\sqrt{\phi_{k,p}}(\phi_{k,l_{1}},\phi_{k,l_{2}},\dots,\phi_{k,p}-1,\dots,\phi_{k,l_{\text{dim}(\boldsymbol{l})}}),
\end{equation}
so the number operator can be defined 
\begin{equation}
\hat{N}_{l}=\hat{a}_{l}^{\dag}\hat{a}_{l}.\label{eq:nl}
\end{equation}
The quantum wavefunctions are written
\begin{equation}
|\psi(t)\rangle=\sum_{k}\psi_{k}(t)\boldsymbol{\phi}_{k},\label{eq:psi_def}
\end{equation}
where $\boldsymbol{\phi}_{k}\in\Phi_{N}$. 

The dimension of this space grows combinatorially fast with increasing
quanta:
\begin{equation}
\text{dim}(\Phi_{N})=\binom{\text{dim}(\boldsymbol{l})+N_{}-1}{N_{}},\label{eq:f_space_size}
\end{equation}
but the second quantized Hamiltonian 
\begin{equation}
\hat{H}=\frac{\delta}{2}\sum_{l\in\boldsymbol{l}}\hat{a}_{l}^{\dag}\hat{a}_{l}(2\pi l)^{2}+\frac{1}{8\pi^{2}\delta}\sum_{p,r,l\in\boldsymbol{l};r\not=l}(r-l)^{-2}\hat{a}_{l}^{\dag}\hat{a}_{p}^{\dag}\hat{a}_{l-r+p}\hat{a}_{r},\label{eq:quant_ham}
\end{equation}
is very sparse. For a mode set $\boldsymbol{l}=\{-n,-n+1,\dots,n\}$,
the sparsity of $\hat{H}$, $s(n)$, only grows as a cubic function
of the maximum absolute mode number: 
\begin{equation}
s(n)=1-\frac{1}{3}n+n^{2}+\frac{4}{3}n^{3}.
\end{equation}
Since $\text{dim}(\boldsymbol{l})=2n+1$, the maximum fraction of
non-zero elements of $\hat{H}$ is
\begin{equation}
\left(\frac{s(n)}{\text{dim}(\Phi_{N})}\right)^{2}\simeq\left(\frac{n^{3}N!(2n)!}{(2n+N)!}\right)^{2}
\end{equation}
when $n\gg1$. This rapidly approaches zero when increasing either
$N$ or $n$, which is ideal for quantum Hamiltonian algorithms. 

For correspondence between the pre-quantized and quantized systems,
the expected value of each number operator
\begin{equation}
\langle\hat{N}_{l}\rangle=\sum_{k}\psi_{k}^{*}\psi_{k}\boldsymbol{\phi}_{k,l}
\end{equation}
should correspond those of each action $I_{l}=a_{l}^{*}a_{l}$ and
the first derivative of each number operator 
\begin{equation}
\partial_{t}\langle\hat{N}_{l}\rangle=\delta(2\pi)^{2}l^{2}\langle\hat{N}_{l}\rangle+\frac{1}{(2\pi)^{2}\delta}\sum_{p,r\in\boldsymbol{l},r\not=l}(r-l)^{2}\langle\hat{a}_{l}^{\dag}\hat{a}_{p}^{\dag}\hat{a}_{l-r+p}\hat{a}_{r}+c.c.\rangle,\label{eq:nl_dot}
\end{equation}
where $c.c.$ denotes the complex conjugate of the preceding term,
should correspond with the first derivative of each action
\begin{equation}
\partial_{t}I_{l}=2\text{Re}\left\{ \frac{\delta}{2}(2\pi)^{2}l^{2}I_{l}+\frac{1}{(2\pi)^{2}\delta}\sum_{p,r\in\boldsymbol{l},r\not=l}(r-l)^{2}a_{l}^{*}a_{p}^{*}a_{l-r+p}a_{r}\right\} .\label{eq:il_dot}
\end{equation}
Because the classical actions add to one, $\sum_{l}I_{l}=1$, while
the quantum number operators add to the total number of quanta, $\sum_{l}\langle\hat{N}_{l}\rangle=N$,
the quantum Eqs. (\ref{eq:nl}) and (\ref{eq:nl_dot}) need to be
made scale invariant. Transform $\langle\hat{N}_{l}\rangle\rightarrow N\langle\hat{N}_{l}\rangle$,
$\delta\rightarrow\sqrt{N}\delta$, and $t\rightarrow t/\sqrt{N}$.
Unless otherwise specified, the scale invariant quantities will be
used from here. For an initial state described by Eq. (\ref{eq:psi_def})
let
\begin{equation}
\psi_{k}(0)=\mathcal{N}\prod_{l\in\boldsymbol{l}}\left\{ e^{-(\mu_{l}-\boldsymbol{\phi}_{k,l})^{2}/\sigma_{l}}e^{i\theta_{Ql}\boldsymbol{\phi}_{k,l}}\right\} ,\label{eq:psi_0}
\end{equation}
with $\mathcal{N}$ a normalization such that $\sum_{k}\psi_{k}^{*}\psi_{k}=1$,
$\mu_{l}$ the mean number of quanta in mode $l$, $\sigma_{l}$ the
standard deviation of quanta in mode $l$, and $\theta_{Ql}$ the
phase difference accumulated while increasing quanta in mode $l$.
For correspondence, choose $\mu_{l}=NI_{l}$ and $\theta_{Ql}=\text{arg}(a_{l})$.
An explanation of these choices may be found in Appendix I. The standard
deviation should be chosen to minimize the uncertainty of both $\langle\hat{N}_{l}\rangle$
and $\partial_{t}\langle\hat{N}_{l}\rangle$ over the the simulation
time. For the purposes of this study, it will be chosen \textit{ad
hoc. }

\section{Simulation}

For a symmetric set of three modes $\boldsymbol{l}=\{-l,0,l\}$, the
equations of motion for the pre-quantized and quantized systems are
simple enough to write explicitly. In the pre-quantized system, the
$-l$ mode evolves according to 
\begin{equation}
\partial_{t}I_{-l}=\delta(2\pi)^{2}l^{2}I_{-l}+\frac{1}{\delta(2\pi)^{2}l^{2}}\left(2I_{0}I_{-l}+a_{-l}^{*}a_{l}^{*}a_{0}a_{0}+a_{0}^{*}a_{0}^{*}a_{-l}a_{l}+\frac{1}{2}I_{-l}I_{l}\right).
\end{equation}
This may be directly compared to the second-quantized version: 
\begin{equation}
\partial_{t}\langle\hat{N}_{-l}\rangle=\delta(2\pi)^{2}l^{2}\langle\hat{N}_{-l}\rangle+\frac{1}{\delta(2\pi)^{2}l^{2}}\left\langle 2\hat{N}_{0}\hat{N}_{-l}+\hat{a}_{-l}^{\dag}\hat{a}_{l}^{\dag}\hat{a}_{0}\hat{a}_{0}+\hat{a}_{0}^{\dag}\hat{a}_{0}^{\dag}\hat{a}_{-l}\hat{a}_{l}+\frac{1}{2}\hat{N}_{-l}\hat{N}_{l}\right\rangle ,\label{eq:quant_number}
\end{equation}
where the Hamiltonian acts to couple an arbitrary basis vector $\{c_{1},c_{2},c_{3}\}\in\Phi_{c_{1}+c_{2}+c_{3}}$
to the basis vectors $\{c_{1}\pm1,c_{2}\mp2,c_{3}\pm1\}$. Thus, despite
the Fock space for three modes having
\begin{equation}
d_{3}=\binom{2+N_{}}{N_{}}=\frac{1}{2}(1+N_{})(2+N_{})
\end{equation}
elements, the space is split into many invariant sub-spaces under
the action of the Hamiltonian. In general, the Hamiltonian of $\text{dim}(\boldsymbol{l})$
will split the Fock space into invariant subspaces of $\text{dim}(\boldsymbol{l})-2$
because it conserves the total quantum number $N$ and does not allow
for single quantum exchanges. Recalling the definition of $|\psi(t)\rangle$
in Eq. (\ref{eq:psi_def}), an arbitrary vector within each subspace
may be written 
\begin{equation}
|\psi(t)\rangle=\sum_{k=0}^{c_{2}/2}\psi_{k}(t)\boldsymbol{\phi}_{k},
\end{equation}
where the basis vectors 
\begin{equation}
\boldsymbol{\phi}_{k}=\{c_{1}+k,c_{2}-2k,c_{3}+k\},
\end{equation}
with $c_{1}$ and $c_{3}$ arbitrary constants and $c_{2}=N-c_{1}-c_{3}$.
For the simplest symmetric case, $c_{1}=c_{3}=0,$ so $c_{2}=N$.
If the initial condition is low-amplitude, and the wavefunction is
expected to remain low-amplitude for all times, i.e. $\sum_{k>k_{max}}|\psi_{k}|^{2}<\varepsilon$
for some $k_{max}$ and $\varepsilon$, then the Hilbert space may
be further truncated to some dimension $d_{3}^{\prime}=1+c_{2}/2<1+N/2$. 

We will now compare pre-quantized and quantized solutions of the Schr\"odinger--Poisson
equations for three modes. An initial condition for the Schr\"odinger--Poisson
equations is chosen, the condition decomposed into modes, a corresponding
quantized initial condition is chosen, and then the amplitudes of
the zero-modes are compared over time. The choice to compare the pre-quantized
wave action $I_{0}$ to the quantized $\langle\hat{N}_{0}\rangle$
is discussed in Appendix II. 

\begin{figure}
\noindent\begin{minipage}[t]{1\columnwidth}%
\includegraphics[width=0.85\linewidth]{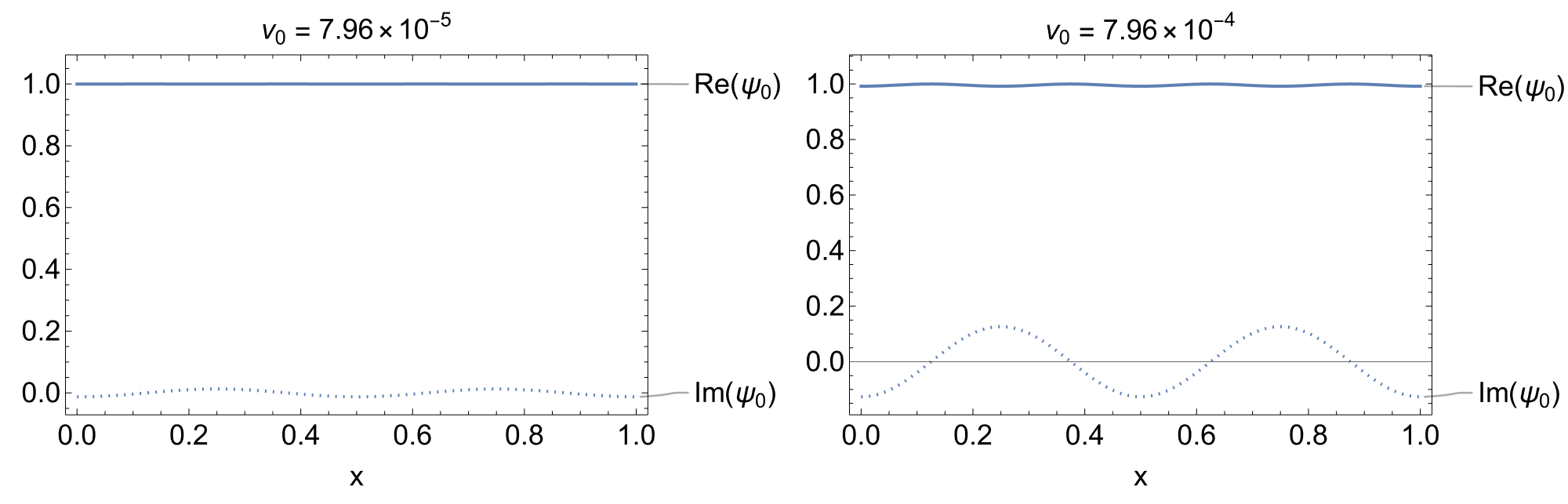}%
\end{minipage}\caption{Real and imaginary parts of Eq. (\ref{eq:cold_wave}) for $\delta=0.0005$,
$k_{0}=4\pi$, and with $v_{0}=1/(400\pi)=7.96\times10^{-5}$ (left)
and $v_{0}=1/(40\pi)=7.96\times10^{-4}$ (right). \protect\label{fig:init_cons}}

\end{figure}

\begin{figure}
\noindent\begin{minipage}[t]{1\columnwidth}%
\includegraphics[width=0.85\linewidth]{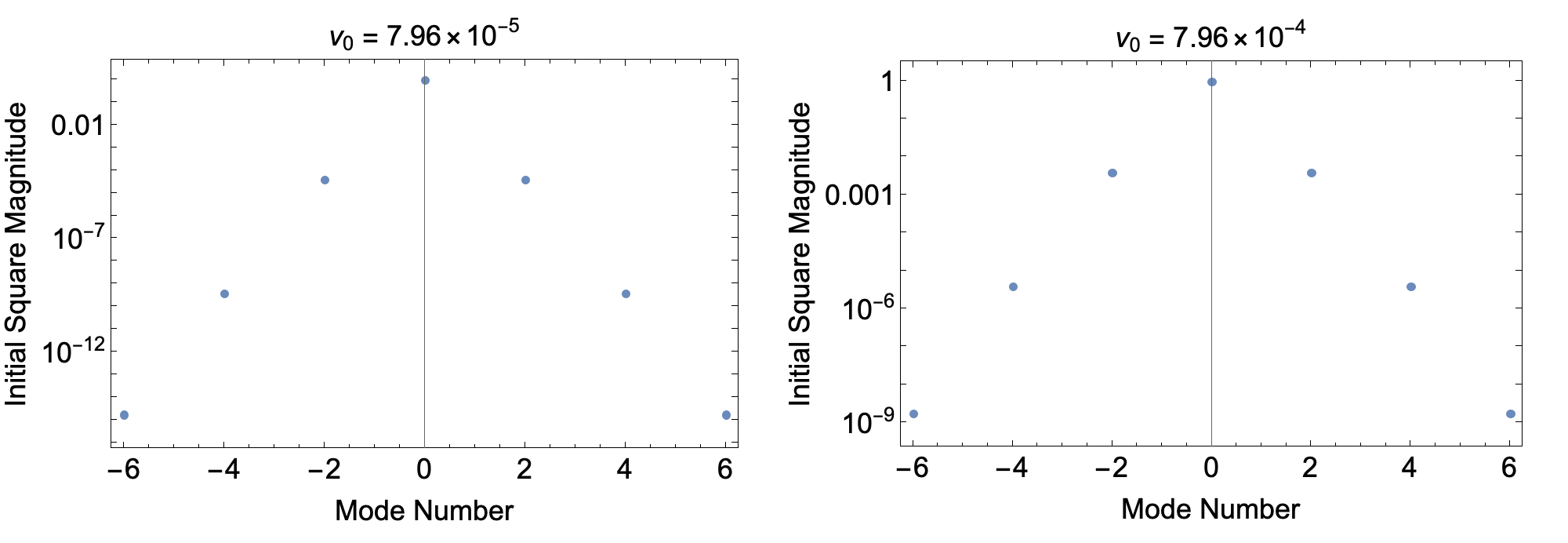}%
\end{minipage}\caption{Square amplitude of modes composing the initial conditions shown in
Fig. \ref{fig:init_cons}. Only the $\boldsymbol{l}\in(-6,6)$ modes
are shown, and odd modes do not appear since they have amplitude 0.
\protect\label{fig:init_modes}}
\end{figure}

Consider an initial condition derived from that found by Bertrand
\textit{et al. }\citep{bertrand80}: 
\begin{equation}
\psi_{p}(x,0)=e^{-i\frac{v_{0}}{\delta k_{0}}\text{cos}(k_{0}x)},\label{eq:cold_wave}
\end{equation}
which corresponds to an initial cold plasma wave $f(x,v,0)=n_{0}\delta(v-v_{0})$.
The square amplitude of the wave is initially spatially homogenous,
so the real and imaginary parts of $\psi_{p}(x,0)$ are shown in Fig.
\ref{fig:init_cons} for two different values of the initial velocity
$v_{0}$ with $k_{0}=4\pi$ and quantum parameter $\delta=0.0005$.
With these initial conditions, $\psi_{p}$ oscillates sinusoidally
with a period of the scaled plasma frequency, $\tau=2\pi$, so $|\psi_{p}|^{2}$
oscillates with a period of $\tau=\pi$. The square amplitudes of
the modes composing these initial conditions, found according to the
mode decomposition in Eq. (\ref{eq:mode_decomp}), are shown in Fig.
\ref{fig:init_modes} and also oscillate with periods of $\tau=\pi$.

With a finite number of modes, the system does not necessarily oscillate
at the plasma frequency, though. In the three mode system with $\boldsymbol{l}=\{-l,0,l\}$,
consider a small perturbation around the pre-quantized equilibrium
$a_{0}=1$. We can determine four governing equations from Eq. (\ref{eq:adot}):
\begin{eqnarray*}
\delta_{t}a_{l} & = & \frac{\delta}{2}a_{l}(2\pi l)^{2}+\frac{1}{\delta(2\pi l)^{2}}(a_{l}+a_{-l}^{*}),
\end{eqnarray*}
and similar equations for $a_{-l}$, $a_{l}^{*}$, and $a_{-l}^{*}$.
This linearization is only valid when $|a_{l}|^{2}\ll4\pi^{4}\delta^{2}l^{4}$,
and the normal modes of this system are $\omega_{\pm}=\pm\sqrt{1+(2\delta\pi^{2}l^{2})^{2}}$
\citep{qin_may_25}. This analysis distinguishes a small amplitude
regime where the system oscillates at the plasma frequency. In the
large amplitude regime, where the linearization is no longer valid,
the system may only oscillate at the plasma frequency via nonlinear
interactions with higher order modes. Denote these two oscillations
the linear plasma oscillation and the nonlinear plasma oscillation.

\begin{figure}
\subfloat[Pre-quantized initial condition with $v_{0}=1/(400\pi)=7.96\times10^{-5}$
. The three modes have initial values $a_{-2,0}=\{-1.78\times10^{-4}-4.5\times10^{-3}i,0.99998-4.5\times10^{-7}i\}$.
The quantum initial condition takes total quanta $N=4\times10^{7}$,
truncated dimension $d_{3}^{\prime}=3001$, mean quanta in each mode
$\mu_{-2,0}=N|a_{-2,0}|^{2}$, accumulated phase differences $\theta_{Q-2,0}=\text{arg}(a_{-2,0})$,
and variances $\sigma_{-2,0}=\{4.0\times10^{6},8.0\times10^{6}\}$. ]{%
\noindent\begin{minipage}[t]{1\columnwidth}%
\includegraphics[width=0.45\linewidth]{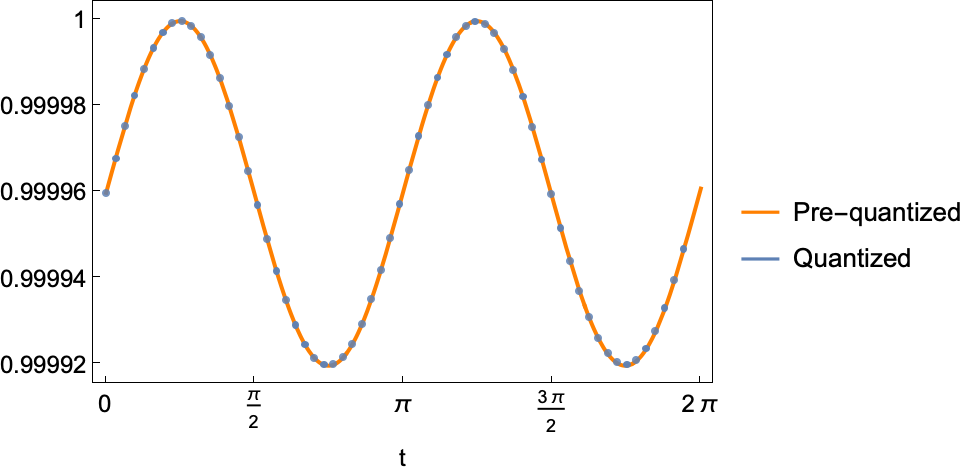}%
\end{minipage}

}\hfill{}\subfloat[Pre-quantized initial condition with $v_{0}=1/(400\pi)=7.96\times10^{-4}$
. The three modes have initial values $a_{-2,0}=\{-0.0019-0.044i,0.998-5\times10^{-5}i\}$.
The quantum initial condition takes total quanta $N=3\times10^{5}$,
truncated dimension $d_{3}^{\prime}=2251$, mean quanta in each mode
$\mu_{-2,0}=N|a_{-2,0}|^{2}$, accumulated phase differences $\theta_{Q-2,0}=\text{arg}(a_{-2,0})$,
and variances $\sigma_{-2,0}=\{2.1\times10^{6},4.2\times10^{6}\}$. ]{%
\noindent\begin{minipage}[t]{1\columnwidth}%
\includegraphics[viewport=0bp 0bp 486bp 235bp,width=0.45\linewidth]{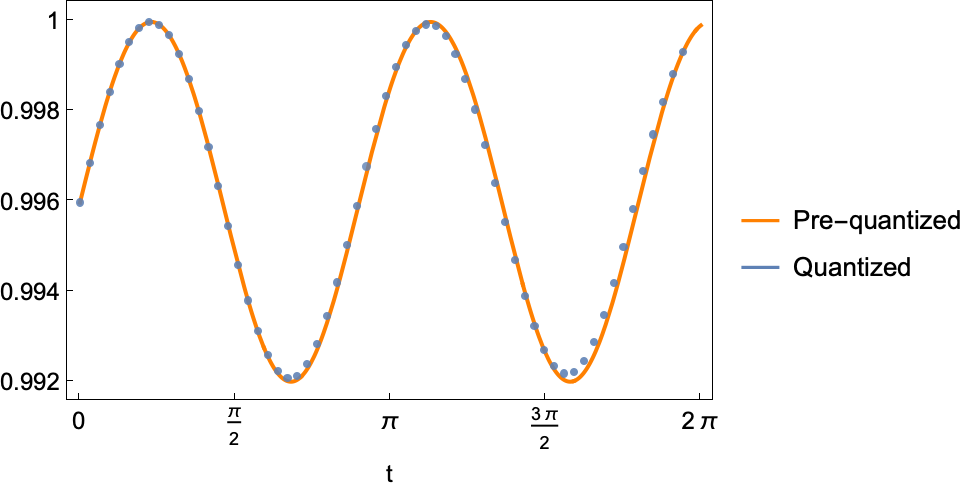}%
\end{minipage}

}

\caption{Evolution of $I_{0}$ and $\langle\hat{N}_{0}\rangle$ according to
the three-mode pre-quantized and quantized Schr\"odinger--Poisson equations
with initial conditions in the linear (a) and nonlinear (b) regimes.
\protect\label{fig:3_wave_comp}}
\end{figure}

Shown in Fig. \ref{fig:3_wave_comp} are the evolution of the zero
mode of the pre-quantized and quantized Schr\"odinger--Poisson equations
in the three mode system with $\boldsymbol{l}=\{-2,0,2\}$ and quantum
parameter $\delta=0.0005$. The quantized initial conditions are determined
according to Eq. (\ref{eq:psi_0}). 
For the $l=\pm2$ modes, the linear plasma oscillation requires $|a_{\pm2}|^{2}\ll.0015$,
so the lower amplitude oscillation of Fig. \ref{fig:3_wave_comp}
a. is a linear plasma oscillation with the expected period of $\pi$.
The higher amplitude oscillation of Fig. \ref{fig:3_wave_comp} b.
is of high enough amplitude that it is non-negligibly nonlinearly coupled to higher order modes, placing it in the nonlinear plasma oscillation regime. In
this truncated system, it oscillates with an erroneously smaller period.
However, in either case the pre-quantized and quantized systems agree
in both the amplitude and frequency of the oscillation.

At least five modes are necessary to capture the nonlinear interaction
between pairs of modes, but the five mode system poses a more significant
challenge to classical simulation than the three mode system. As in
the three mode system, the action of the Hamiltonian divides the five
mode system into many invariant subspaces. An arbitrary vector within
each subspace may be written 
\begin{equation}
|\psi(t)\rangle=\sum_{j=0}^{c_{3}/2}\sum_{k=0}^{(c_{3}-2j)/2}\sum_{m=-\text{min}(j,k)}^{\text{min}(j,c_{3}-2j-2k)}\psi_{j,k,m}(t)\boldsymbol{\phi}_{j,k,m},
\end{equation}
where
\begin{equation}
\boldsymbol{\phi}_{j,k,m}=\{c_{1}+j,c_{2}+k+m,c_{3}-2j+2k-m,c_{4}-k-m+k,c_{5}+j+m\}.
\end{equation}
The total number of quanta is $N=\sum_{l}c_{l}$. In order to have
the same resolution in each of the three dimensions as used in the
simulation of Fig. \ref{fig:3_wave_comp} b., the five mode system
would need a truncated dimension $d_{5}^{\prime}\simeq d_{3}^{\prime3}>10^{10}.$
The problem is more severe when considering the decades of magnitude
difference between subsequent modes given the initial condition, Eq.
(\ref{eq:cold_wave}), evident in Fig. \ref{fig:init_modes}. In that
case, the $l\pm2$ modes oscillate with amplitudes $10^{3}$ times
larger than the $l\pm4$ modes, which would require $d_{5}^{\prime}\simeq10^{13}$.
Even with the Hamiltonian being sparse, this would be unfeasible to
simulate on a classical computer, so let us only consider initial
conditions for which the relative dynamic range between modes is small. 

\begin{figure}
\includegraphics[scale=0.5]{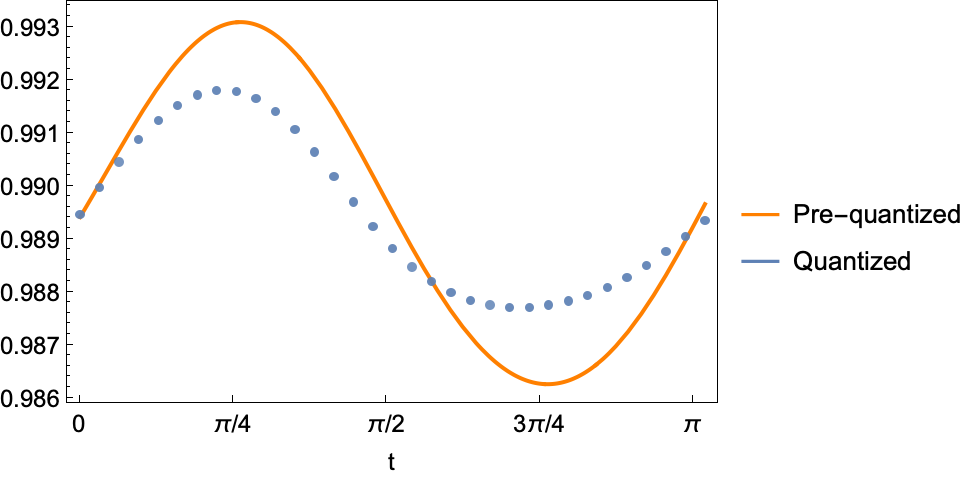}\caption{Evolution of $I_{0}$ and $\langle\hat{N}_{0}\rangle$ according to
the five-mode pre-quantized and quantized Schr\"odinger--Poisson equations,
respectively. The pre-quantized initial condition is symmetric with
$a_{-4,-2,0}=\{-9.51\times10^{-3}-3.57\times10^{-2}i,3.21\times10^{-4}-6.43\times10^{-2}i,0.9945-7.97\times10^{-4}i\}$.
The quantum initial condition has total quanta $N=2.43\times10^{5}$,
a truncated number of basis vectors $d_{5}^{\prime}\simeq1.5\times10^{6}\simeq j_{max}\times k_{max}\times35$,
mean initial quanta in each mode $\mu_{-4,-2,0}=N|a_{-4,-2,0}|^{2}$,
and accumulated phase differences $\theta_{Q-4,-2,0}=\text{arg}(a_{-4,-2,0})$.}

\label{fig:5mode}
\end{figure}

The results of two such simulations featuring nonlinearities are shown
below. Each uses the mode set $\boldsymbol{l}=\{-4,-2,0,2,4\}$. For
the first simulation, shown in Fig. \ref{fig:5mode}, a nonlinear
plasma oscillation with $c_{3}=N=2.43\times10^{5}$ quanta is captured.
The basis vectors $\{\phi_{j,k,m}\}$ are truncated so that so that
$j\le j_{max}=300$, $k\le k_{max}=150$, and $-17\le m\le17$. Because
the initial condition is symmetric, i.e. $a_{-2}=a_{2}$ and $a_{-4}=a_{4}$,
the antisymmetric dimension of the basis vectors $m$ is taken to
be relatively narrow around zero. The amplitude of the oscillation
is approximately the same as that of the three-mode nonlinear case
in Fig. \ref{fig:3_wave_comp} b., placing it within the nonlinear
regime, yet the inclusion of the additional $l=\pm4$ modes allows
both the pre-quantized and quantized solutions to capture the expected
plasma frequency. The deviation of the quantized trajectory and the
the pre-quantized Schr\"odinger--Poisson solutions comes from the wavefunction
being poorly resolved when it is squeezed at both the maximum and
minimum of its trajectory. Increasing the total number of quanta,
and $j_{max}$, $k_{max}$, and $m_{max}$ proportionally, will reduce
the error. 

\begin{figure}
\includegraphics[scale=0.5]{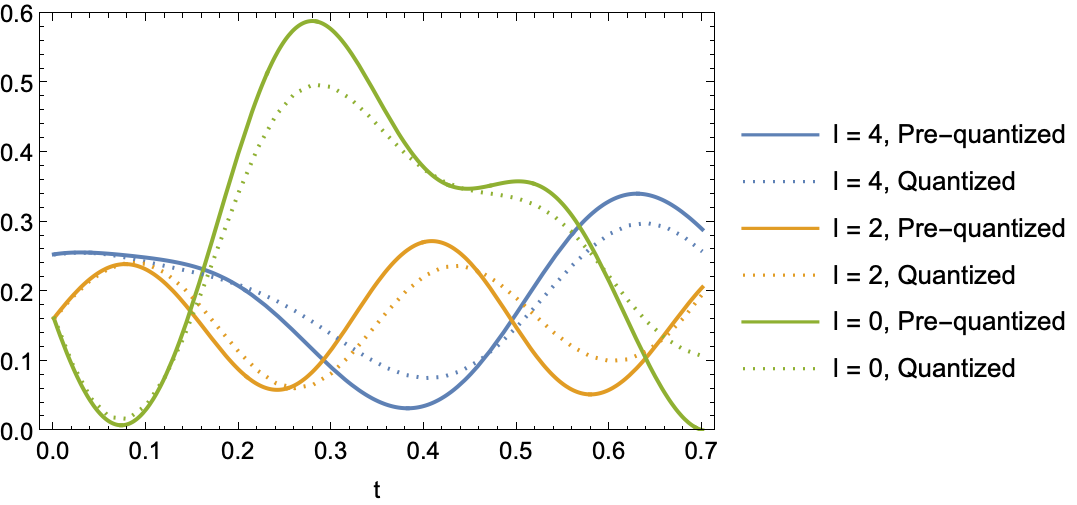}\caption{Evolution of $I_{0}$, $I_{2}$, $I_{4}$, $\langle\hat{N}_{0}\rangle$,
$\langle\hat{N}_{2}\rangle$, and $\langle\hat{N}_{4}\rangle$ according
to the five-mode pre-quantized and quantized Schr\"odinger--Poisson
equations, respectively. The pre-quantized initial condition is symmetric
with $a_{-4,-2,0}=\{0.36+0.36i,0.29+0.29i,-0.29+0.29i\}$. The quantum
initial condition has total quanta $N=400$, mean initial quanta in
each mode $\mu_{-4,-2,0}=N|a_{-4,-2,0}|^{2}$, and accumulated phase
differences $\theta_{Q-4,-2,0}=\text{arg}(a_{-4,-2,0})$.}

\label{fig:5mode_nonlin}
\end{figure}

The second simulation includes no truncation of the basis vectors,
but the total number of quanta is much smaller: $N=400$. This gives
a set of $1.49\times10^{6}$ basis vectors. Because the full state
space is captured, it is possible to simulate much larger amplitudes,
shown in Fig. \ref{fig:5mode_nonlin}. As with the previous simulation,
proximity to the maxima and minima, where the wavefunction becomes
poorly resolved, increases the deviation of the simulation from the
pre-quantized Schr\"odinger--Poisson solution. 

\section{Discussion}

The nonlinear plasma oscillation can be captured in the Schr\"odinger--Poisson
system linearized by second-quantization. Via a Wigner transformation,
Eq. \ref{eq:wig}, this solution can then be used to model a solution
of the nonlinear Vlasov--Poisson equations. This method may provide
future, fault-tolerant quantum computers with the means to out-perform
classical computers in simulations of the Vlasov--Poisson system. 

Setting aside the issues of state preparation and the choice of observables
to measure, the integration time of the second-quantized Schr\"odinger--Poisson
system on a quantum computer may scale very favorably with the system
size, assuming that one chooses to integrate a probability distribution
rather than a single trajectory. Consider that the total number of
quanta in the system will be proportional to the number of modes simulated
$N\propto\text{dim}(\boldsymbol{l})$, and the condition number of
the Hamiltonian will scale $\kappa\propto N$. The sparsity of the
system $s\propto\text{dim}(\boldsymbol{l})^{3}$. The integration
time using an efficient quantum algorithm for Hamiltonian simulation
\citep{low17} will then be $\tau_{simQ}\propto\kappa s\propto\text{dim}(\boldsymbol{l})^{4}$.
This is, of course, much longer than than that of a classical algorithm
operating on the nonlinear system which, owing to the double sum in
the Hamiltonian, will be at worst $\tau_{simC}\propto\text{dim}(\boldsymbol{l})^{2}$.
However, because the quantum system evolves a probability distribution
rather than a single trajectory, one can initialize multiple classical
initial conditions and linearly evolve them simultaneously. 

Suppose a quantum probability distribution corresponding to a classical
initial condition has a characteristic width along a particular mode
direction, $w_{l}$, and the basis vectors have a quantum resolution,
the number of quanta a mode can vary by, $R_{l}$. The width, $w_{l}$,
will be set by the variance $\sigma_{l}$ in Eq. (\ref{eq:psi_0}),
and $R_{l}$ will be $N/2$, unless the $l$ dimension is truncated.
The fraction of the total phase space volume taken up by this particular
wave packet is then $\prod_{l\in\boldsymbol{l}}w_{l}/R_{l}$. Assuming
$w_{l}$ and $R_{l}$ to be relatively constant for most modes, $w_{l}/R_{l}\sim w/R$,
for some characteristic $w$ and $R$. If the wave packet's size does
not grow significantly over the simulation time, one can fill the
remaining quantum phase space volume with wave packets initialized
with varying probabilities. Thus, the effective quantum integration
time is divided by the total number of wave packets one may simultaneously
simulate:
\begin{equation}
\tau_{eff}\propto\text{dim}(\boldsymbol{l})^{4}\left(\frac{w}{R}\right)^{\dim(\boldsymbol{l})}.
\end{equation}
For large $\text{\text{dim}}(\boldsymbol{l})$, $\tau_{eff}\ll\tau_{simC}$,
regardless of the size of $w/R$. Future work will need determine
exact size of physically relevant simulations for more accurate simulation
times to be determined. 

In conclusion, second quantization provides an effective means of
linearizing physically interesting nonlinear differential equations.
We have detailed the second quantization of the Vlasov--Poisson system
and shown how it can accurately capture a nonlinear plasma oscillation.
Although quantum computers will never be able to efficiently compute
time series, as shown in Figs. \ref{fig:3_wave_comp} and \ref{fig:5mode},
for specific observables measured at the end of some integration,
quantum Hamiltonian simulations can be much more efficient than classical
algorithms. For the Vlasov--Poisson system, this linearization may
provide future quantum computers with the means to simulate nonlinear
astrophysical and plasma phenomena much more efficiently than previously
thought possible. 
\begin{acknowledgments}
This research is supported by the U.S. Department of Energy (DE-AC02-09CH11466). 
\end{acknowledgments}

\section*{Appendix I}

We will show if $\mu_{l}=NI_{l}$ and $\theta_{Ql}=\text{arg}(a_{l})$
in Eq. (\ref{eq:psi_0}), 
\[
\psi_{k}(0)=\mathcal{N}\prod_{l\in\boldsymbol{l}}\left\{ e^{-(\mu_{l}-\phi_{k,l})^{2}/\sigma_{l}}e^{i\theta_{Ql}\phi_{k,l}}\right\} ,
\]
the classical and quantum equations of motion will be in correspondence,
i.e. $\langle\hat{N}_{l}\rangle\simeq I_{i}$ and $\partial_{t}\langle\hat{N}_{l}\rangle\simeq\partial_{t}I_{l}$.
The first relation is obvious from direct calculation. If the mean
occupation number for a particular mode $\mu_{l}\in\mathbb{N}$, then
\begin{eqnarray}
\langle\hat{N}_{l}\rangle & = & \sum_{k}\psi_{k}^{*}\psi_{k}\phi_{k,l}\\
 & = & \mathcal{N}^{2}\sum_{k}\left\{ \left(\prod_{r\in\boldsymbol{l}}e^{-2(\mu_{r}-\phi_{k,r})^{2}/\sigma_{r}}\right)\phi_{k,l}\right\} \\
 & = & \mathcal{N}^{2}\sum_{k}\left\{ \left(\prod_{r\in\boldsymbol{l},r\not=l}e^{-2(\mu_{r}-\phi_{k,r})^{2}/\sigma_{r}}\right)e^{-2(\mu_{l}-\phi_{k,l})^{2}/\sigma_{l}}\phi_{k,l}\right\} .
\end{eqnarray}
The first and second equations are just substitutions of the definitions
for $\langle\hat{N}_{l}\rangle$ and $\psi_{k}(0)$. In the third
equation, the product is expanded. Next, $\phi_{k,l}$ will be written
explicitly. The sum will be split into two parts, one where $k$ takes
all such values where $\phi_{k,l}=\mu_{l}$ and one which has the
other terms. This gives: 
\begin{eqnarray}
\langle\hat{N}_{l}\rangle & \simeq & \mathcal{N}^{2}\sum_{p>0}\sum_{k\ni\phi_{k,l}=\mu_{l}+p}\left\{ \left(\prod_{r\in\boldsymbol{l},r\not=l}e^{-2(\mu_{r}-\phi_{k,r})^{2}/\sigma_{r}}\right)e^{-2p^{2}/\sigma_{l}}(\mu_{l}+p+\mu_{l}-p)\right\} \\
 &  & +\mathcal{N}^{2}\sum_{k\ni\phi_{k,l}=\mu_{l}}\left\{ \left(\prod_{r\in\boldsymbol{l},r\not=l}e^{-2(\mu_{r}-\phi_{k,r})^{2}/\sigma_{r}}\right)\mu_{l}\right\} \\
 & = & \mu_{l}\mathcal{N}^{2}\sum_{k}\left\{ \prod_{r\in\boldsymbol{l}}e^{-2(\mu_{r}-\phi_{k,r})^{2}/\sigma_{r}}\right\} \\
 & = & \mu_{l}.
\end{eqnarray}
The equality is only approximate in the first line because the initial
condition may not be centered in the space of basis vectors. In the
most extreme case, where $\mu_{l}=0$, Eq. (\ref{eq:psi_0}) will
not allow for quantum--classical correspondence. However, for any
other value of $\mu_{l}$, in the classical limit where the total
number of quanta $N\rightarrow\infty$ and $\sigma\ll N^{2}$, this
symmetry condition may be ignored. This proof can be easily extended
to the case where $\mu_{l}$ is any positive rational number, rather
than an integer. 

The second correspondence relation can be determined with Eqs. (\ref{eq:nl_dot})
and (\ref{eq:il_dot}). Writing
\begin{eqnarray*}
\partial_{t}\langle\hat{N}_{l}\rangle & = & \delta(2\pi)^{2}l^{2}\langle\hat{N}_{l}\rangle+\frac{1}{(2\pi)^{2}\delta}\sum_{p,r\in\boldsymbol{l},r\not=l}(r-l)^{2}\langle\hat{a}_{l}^{\dag}\hat{a}_{p}^{\dag}\hat{a}_{l-r+p}\hat{a}_{r}+c.c.\rangle\\
 & = & \dot{N}_{l1}+\dot{N}_{l2},
\end{eqnarray*}
and 
\begin{eqnarray*}
\partial_{t}I_{l} & = & 2\text{Re}\left\{ \frac{\delta}{2}(2\pi)^{2}l^{2}I_{l}+\frac{1}{(2\pi)^{2}\delta}\sum_{p,r\in\boldsymbol{l},r\not=l}(r-l)^{2}a_{l}^{*}a_{p}^{*}a_{l-r+p}a_{r}\right\} \\
 & = & \dot{I}_{l1}+\dot{I}_{l2},
\end{eqnarray*}
it is clear that the first correspondence condition, $\mu_{l}=NI_{l}$,
gives 
\begin{eqnarray*}
\dot{N}_{l1} & = & \delta(2\pi)^{2}l^{2}\langle\hat{N}_{l}\rangle\\
 & \simeq & \delta(2\pi)^{2}l^{2}I_{l}\\
 & = & \dot{I}_{l1}.
\end{eqnarray*}
To show $\dot{N}_{l2}\simeq\dot{I}_{l2}$, one needs to show $\langle\hat{a}_{l}^{\dag}\hat{a}_{p}^{\dag}\hat{a}_{l-r+p}\hat{a}_{r}+c.c.\rangle\simeq a_{l}^{*}a_{p}^{*}a_{l-r+p}a_{r}+c.c.$
for all $l$, $p$, and $r\not=l$. As an example, we will show this
for the five mode system $\boldsymbol{l}=\{-4,-2,0,2,4\}$ with $l=0$,
$p=-2$, and $r=2$. Writing the classical amplitudes in polar form,
such that $a_{l}=|a_{l}|e^{i\theta_{l}}$, 
\begin{equation}
a_{0}^{*}a_{-2}^{*}a_{-4}a_{2}+c.c.=|a_{0}||a_{-2}||a_{-4}||a_{2}|\text{cos}(\theta_{-4}+\theta_{2}-\theta_{0}-\theta_{-2}).
\end{equation}
Beginning with the quantum state $\Phi=\sum_{j,k}\psi_{j,k}\boldsymbol{\phi}_{j,k}$,
where the basis vectors
\begin{equation}
\boldsymbol{\phi}_{j,k}=\{c_{-4}-k,c_{-2}-j+2k,c_{0}-k+2j,c_{2}-j,c_{4}\},
\end{equation}
the quantum $\langle\hat{a}_{0}^{\dag}\hat{a}_{-2}^{\dag}\hat{a}_{-4}\hat{a}_{2}+c.c.\rangle$
may be calculated directly: 
\begin{eqnarray*}
\langle\hat{a}_{0}^{\dag}\hat{a}_{-2}^{\dag}\hat{a}_{-4}\hat{a}_{2}+c.c.\rangle & = & \Bigg\langle\sum_{l,m}\psi_{l,m}\boldsymbol{\phi}_{l,m}\Big|\sum_{j,k}\psi_{j,k}\Big\{(c_{0}-k+2j+1)^{1/2}\\
 &  & \times(c_{-2}-j+2k+1)^{1/2}(c_{-4}-k)^{1/2}(c_{2}-j)^{1/2}\boldsymbol{\phi}_{j+1,k+1}\\
 &  & +(c_{0}-k+2j)^{1/2}(c_{-2}-j+2k)^{1/2}(c_{-4}-k+1)^{1/2}\\
 &  & \times(c_{2}-j+1)^{1/2}\boldsymbol{\phi}_{j,k}\Big\}\Bigg\rangle.
\end{eqnarray*}
Next, the sums may be relabeled, and the expectation value calculated:
\begin{eqnarray*}
\langle\hat{a}_{0}^{\dag}\hat{a}_{-2}^{\dag}\hat{a}_{-4}\hat{a}_{2}+c.c.\rangle & = & \sum_{j,k}\big(\psi_{j+1,k+1}^{\dag}\psi_{j,k}(c_{0}-k+2j+1)^{1/2}(c_{-2}-j+2k+1)^{1/2}\\
 &  & \times(c_{-4}-k)^{1/2}(c_{2}-j)^{1/2}+\psi_{j-1,k-1}^{\dag}\psi_{j,k}(c_{0}-k+2j)^{1/2}\\
 &  & \times(c_{-2}-j+2k)^{1/2}(c_{-4}-k+1)^{1/2}(c_{2}-j+1)^{1/2}\big)\\
 & \simeq & \sum_{j,k}2\text{Re}(\psi_{j+1,k+1}^{\dag}\psi_{j,k})(c_{0}-k+2j+1)^{1/2}\\
 &  & \times(c_{-2}-j+2k+1)^{1/2}(c_{-4}-k)^{1/2}(c_{2}-j)^{1/2}
\end{eqnarray*}
The final line is true assuming that $\psi_{j,k}$ is negligible near
its boundaries and that differences of a single quantum cannot be
distinguished in the classical limit. Now, take $\{c_{-4},c_{-2},c_{0},c_{2},c_{4}\}=\{|a_{-4}|^{2},|a_{-2}|^{2},|a_{0}|^{2},|a_{2}|^{2},|a_{4}|^{2}\}$,
and $\mu_{l}=Nc_{l}$ for all $l$. In the classical limit, $\psi_{j,k}$
will only be nonzero in a narrow band around $k=j=0$, so $\langle\hat{a}_{l}^{\dag}\hat{a}_{p}^{\dag}\hat{a}_{l-r+p}\hat{a}_{r}+c.c.\rangle\simeq a_{l}^{*}a_{p}^{*}a_{l-r+p}a_{r}+c.c.$
becomes 
\begin{eqnarray}
\text{cos}(\theta_{-4}+\theta_{2}-\theta_{0}-\theta_{-2}) & \simeq & \sum_{j,k}\text{Re}(\psi_{j+1,k+1}^{\dag}\psi_{j,k})\\
 & \simeq & \sum_{j,k}\left|\psi_{j,k}\right|^{2}\text{cos}\left(\theta_{Q_{-4}}+\theta_{Q_{2}}-\theta_{Q_{0}}-\theta_{Q_{-2}}\right)\\
 & = & \text{cos}\left(\theta_{Q_{-4}}+\theta_{Q_{2}}-\theta_{Q_{0}}-\theta_{Q_{-2}}\right)
\end{eqnarray}
Thus, with the initial condition from Eq. (\ref{eq:psi_0}), $\psi_{k}(0)=\mathcal{N}\prod_{l\in\boldsymbol{l}}\left\{ e^{-(\mu_{l}-\boldsymbol{\phi}_{k,l})^{2}/\sigma_{l}}e^{i\theta_{Ql}\boldsymbol{\phi}_{k,l}}\right\} $,
and taking $\theta_{Ql}=\text{arg}(a_{l})$, we find correspondence
in the classical limit. The argument proceeds the same for any $l$,
$p$, and $r$ and may be generalized to any number of modes. With
$\dot{N}_{l2}\simeq\dot{I}_{l2}$, we have $\partial_{t}\langle\hat{N}_{l}\rangle\simeq\partial_{t}I_{l}$.

\section*{Appendix II}

Throughout this work, prequantized and quantized Schr\"odinger--Poisson
solutions are compared via the pre-quantized wave actions $I_{l}$
and the expectation of the quantum number operators $\langle N_{l}\rangle$.
The square amplitude of the pre-quantized wavefunction may also act
as an operator: $|\psi_{p}(x,t)|^{2}\rightarrow|\hat{\psi}_{p}(x,t)|^{2}$;
however it is not used here because its evaluation complicates the
simulation of the three mode system. After simplification, the expected
value of the pre-quantized square wavefunction operator becomes: 
\begin{equation}
\left\langle \left|\hat{\psi}_{p}(x,t)\right|^{2}\right\rangle =1+\sum_{l\in\boldsymbol{l}}\sum_{r\not=l}e^{2\pi ix(l-r)}\langle a_{l}^{\dag}(t)a_{r}(t)\rangle.
\end{equation}
Unfortunately, this operator won't have spatial or time dependence
when acting on a three-dimensional symmetric subspace. The expression
of $\hat{a}_{l}^{\dag}\hat{a}_{r}$ for $r\not=l$ on a basis vector
$\phi_{k}=\{k,c_{2}-2k,k\}$ won't lie in the original space, e.g.
for the space $\boldsymbol{l}\in\{-l,0,l\}$:
\begin{equation}
\hat{a}_{l}^{\dag}\hat{a}_{0}\phi_{k}\propto\{k,c_{2}-2k-1,k+1\}\not\in\{\phi_{k}\forall k\}.
\end{equation}
Thus, in the restricted space, 
\begin{equation}
\left\langle |\hat{\psi}_{p}(x,t)|^{2}\right\rangle =1.
\end{equation}
One needs to simulate several systems which don't interact through
the Hamiltonian to find a satisfactory corollary to $|\psi_{p}(x,t)|^{2}.$
Going to higher powers of the wavefunction's amplitude will not resolve
this correspondence problem. Indeed, one may find 
\begin{equation}
\left\langle |\hat{\psi}_{p}(x,t)|^{4}\right\rangle =1+2\left(N_{-l}N_{0}+N_{0}N_{l}+N_{-l}N_{l}+\left\langle \hat{a}_{0}^{\dag}\hat{a}_{0}^{\dag}\hat{a}_{-l}\hat{a}_{l}+\hat{a}_{-l}^{\dag}\hat{a}_{l}^{\dag}\hat{a}_{0}\hat{a}_{0}\right\rangle \right),
\end{equation}
has no spatial dependence, though it does depend on time. 
\begin{acknowledgments}
This research was supported by the U.S. Department of Energy (DE-AC02-09CH11466).
\end{acknowledgments}

\bibliographystyle{plain}
\bibliography{big_bib}

\end{document}